\documentclass[aps,prb,twocolumn,superscriptaddress,showpacs]{revtex4}
\usepackage{graphicx}
\begin{document}


\title{ Electronic structure of unidirectional superlattices in
crossed electric and magnetic fields and related  terahertz 
oscillations}
\author{L.\ Smr\v{c}ka} 
\email{smrcka@fzu.cz}
\affiliation{Institute of Physics, Academy of Science of the Czech
Republic, v.v.i.,\\ Cukrovarnick\'{a} 10, 162 53 Praha  6, Czech Republic}
\author{N.\ A.\ Goncharuk}
\affiliation{Institute of Physics, Academy of Science of the Czech
Republic, v.v.i.,\\ Cukrovarnick\'{a} 10, 162 53 Praha  6, Czech Republic}
\author{M.\ Orlita}
\affiliation{Institute of Physics, Academy of Science of the Czech
Republic, v.v.i.,\\ Cukrovarnick\'{a} 10, 162 53 Praha  6, Czech Republic}
\affiliation{Charles University, Faculty of Mathematics and Physics, 
Institute of Physics, Ke Karlovu 5,
121 16 Prague 2, Czech Republic}
\author{R.\ Grill}
\affiliation{Charles University, Faculty of Mathematics and Physics, 
Institute of Physics, Ke Karlovu 5,
121 16 Prague 2, Czech Republic}

\affiliation{}


\date{\today}

\begin{abstract}
We have studied Bloch electrons in a perfect unidirectional
superlattice subject to crossed electric and magnetic fields, where
the magnetic field is oriented ``in-plane'', i.e. in parallel to the sample
plane.  Two orientation of the electric field are considered.  It is
shown that the magnetic field suppresses the intersubband tunneling of
the Zener type, but does not change the frequency of Bloch
oscillations, if the electric field is oriented perpendicularly to
both the sample plane and the magnetic field.  The electric field
applied in-plane (but perpendicularly to the magnetic field) yields
the step-like electron energy spectrum, corresponding to the
magnetic-field-tunable oscillations alternative to the Bloch ones.
\end{abstract}


\maketitle

\section{Introduction\label{Intro}}
Semiconductor structures are considered as a perspective source of
persistent terahertz radiation.~\cite{cap} In semiconductor
superlattices the radiation is generated by Bloch oscillations driven
by the electric field $\vec{\mathcal{E}}$ applied in parallel with the
growth direction, i.e. with the superlattice lattice vector.  Under
influence of the electric field the Wannier-Stark ladder of
quasi-stationary states is formed. The energy of emitted photons is
determined by the separation between neighboring levels of the ladder,
$\hbar\omega_{BO}= |e|\mathcal{E}a$, where $\omega_{BO}$ is the Bloch
oscillation frequency and $a$ denotes the period of the superlattice.
Even though the idea of Bloch oscillations is old, it took a long time
to find their experimental evidence.~\cite{Feldman,Waschke,Deko,
Cho,Lys}  For brief review see, e.g., Hartmann {\it et
al.}~\cite{Hart} and Leo.~\cite{Leo}

The strong transversal magnetic field has been used to quantize the
free in-plane electron motion and to convert the
quasi-three-dimensional electron structure of a superlattice to
quasi-one-dimensional Landau subbands, with the aim to change the
electron dynamics and improve the condition for the terahertz
emission, as described by Patan\`{e} {\it et al.},~\cite{Patane}  
Scalari~{\it et al.},\cite{Scalari} and references therein.   
 
Completely different approach was used in our recent
publication,~\cite{or} where we have theoretically studied the
influence of the strong in-plane magnetic field on the electronic
structure of superlattices.  We have considered the superlattice
subject to crossed electric and magnetic fields, $\vec{\mathcal{E}}$
and $\vec{\mathcal{B}}$, both applied ``in-plane'',
i.e. perpendicularly to the modulation direction, as an alternative
source of radiation.  In that case electrons are driven in parallel
with the lattice vector by the Lorentz force. As electrons tunnel
through the barriers, the cyclic motion along the electric field
direction is superimposed to their otherwise straight-line drift due
to the Lorentz force. The corresponding terahertz frequency
$\omega_\mathcal{B_{\parallel}}$ is related to this cyclic motion and
depends not only on the electric field $\vec{\mathcal{E}}$, as in the
case of Bloch oscillations, but also on the applied magnetic field
$\vec{\mathcal{B}}$. The frequency is given by
$\omega_\mathcal{B_{\parallel}}= 2\pi v_d/a$ where $v_d=\mathcal{E}/
\mathcal{B}$ is the electron drift velocity.

Terahertz oscillations in still another configuration of the crossed
fields was investigated experimentally by Qureshi,~\cite{qureshi}
who employed the transversal electric field, as in the standard Bloch
configuration, combined with a strong in-plane magnetic field. The
chaotic dynamics of electrons in the presence of the crossed fields
and in the tilted magnetic fields was theoretically studied in
papers.~\cite{Wong, From1,From2}
\begin{figure}[b]
\begin{center}
\includegraphics[width=\linewidth]{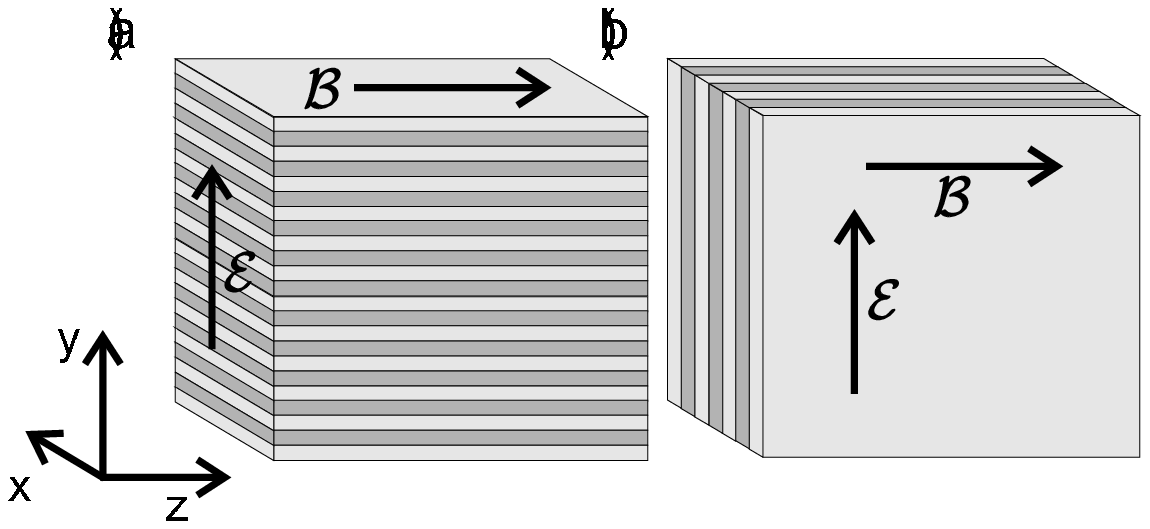}
\caption{\label{fig1}The orientations of the electric and magnetic field with
respect to the superlattice structure. Configurations (a) and (b) are
considered in sections III. and IV., respectively.}
\end{center}
\end{figure}

For Bloch oscillations, the very important issue is the intersubband
tunneling of the Zener type. Lot of efforts were spent to clarify the
condition under which Bloch oscillations could be observed and this
problem is not yet definitely solved.~\cite{ne,so,ya06,ya05} The
same question needs to be addressed to the alternative
magnetic-field-induced oscillations as we did not pay enough attention
to this point in our previous publication.~\cite{or}

We will study the electrons in the presence of crossed magnetic and
electric fields and subject to a unidirectional potential with the
period determined by the lattice vector $\vec{a}$ oriented in the
growth direction of the superlattice.  The considered geometrical
arrangements of the fields and the superlattice are shown in
Fig.~\ref{fig1}.  The magnetic field acts in the direction $z$ with a
constant intensity $\mathcal{B}$ and the electric field of a constant
intensity $\mathcal{E}$ is parallel with the $y$ axis. Two directions
of the lattice vector are considered with respect to the orientation
of $\vec{\mathcal{E}}$: the parallel one, $\vec{a}\equiv (0,a,0)$,
with the periodic potential potential $V(y)=V(y+a)$, and the
perpendicular one, $\vec{a}\equiv (a,0,0)$, with the periodic
potential $V(x)=V(x+a)$. The former configuration corresponds to the
Bloch oscillations, the latter one to the alternative oscillations.

While the three-dimensional superlattices subject to in-plane electric
and magnetic fields have been proposed as an alternative source of the
terahertz oscillations, in our theoretical analysis the simpler
two-dimensional model is employed. As the $z$-dependent part of the
three-dimensional Hamiltonian, $p_z^2/2m$, does not play any role in
the following theoretical consideration, we can omit it for simplicity
without loss of generality.
\section{A free electron in crossed fields\label{Free}}
We  start with description  of a free electron in crossed
fields.~\cite{davies} In general, the two-dimensional
Hamiltonian of an electron in the electromagnetic field reads
\begin{equation}
\label{HAM}
H=\frac{1}{2m}\left(\vec{p}-e\vec{A}\right)^2+e\,\phi,
\end{equation}
where
\begin{equation}
\vec{\mathcal{E}} = - {\rm grad}\,\phi ,\,\,\, \vec{\mathcal{B}} 
= {\rm curl}\, \vec{A},
\end{equation}
and $e=-|e|$. If we further assume $\phi=-\mathcal{E}\,y$ and employ
the Landau gauge of the vector potential
$\vec{A}=(-\mathcal{B}\,y,0,0)$, the Hamiltonian (\ref{HAM}) reduces
to
\begin{equation}
\label{H0}
H_0= \frac{1}{2m}\left(p_x-|e|\mathcal{B}\,y\right)^2+
\frac{p_y^2}{2m} + |e|\mathcal{E}\,y.
\end{equation}
Since $H_0$ commutes with $p_x$, the separation of variables in the
corresponding Schr\"odinger equation is possible.
 
The eigenvalues and eigenstates of $H_0$ are only slightly different
from the more frequently described zero-electric-field case
\begin{equation}
\label{H0tilde}
\widetilde{H}_0= \frac{1}{2m}\left(p_x-|e|\mathcal{B}\,y\right)^2+
\frac{p_y^2}{2m}.
\end{equation}
The eigenfunctions of $\widetilde{H}_0$ can be written in the form
\begin{equation}
\psi_{0nk_x}(\vec{r})=\frac{1}{\sqrt{L_x}} 
{\mathrm e}^{ik_xx}\varphi_{0nk_x}(y),  
\end{equation}
where $k_x=j\, 2\pi/L_x,\,\, j=1,2,\cdots,N$ and the eigenenergies
read
\begin{equation}
\label{Etilde}
\widetilde{E}_{0n}(k_x)=\hbar\omega\left(n+\frac{1}{2}\right), \,\,\
\omega=\frac{|e|\mathcal{B}}{m},\,\, n=0,1,\cdots.
\end{equation}
$L_x$ and $L_y$ denote the sample dimensions.

The functions $\varphi_{0nk_x}(y)$ are the eigenfunctions of the
harmonic oscillator,
\begin{equation}
\label{0phi}
\varphi_{0nk_x}(y)=\frac{1}{\sqrt{2^nn!\sqrt{\pi}\ell}}\,
\exp{\left[-\frac{(y-y_0)^2}{2\ell^2}\right]}
H_n\left(\frac{y-y_0}{\ell}\right),
\end{equation}
where $H_n$ are the Hermit polynomials and $\ell^2=\hbar/m\omega$. The
coordinate $y_0=\ell^2k_x$ relates the center of the cyclotron orbit,
given by $\langle\varphi_{0nk_x}|y|\varphi_{0nk_x}\rangle = y_0$, to the
wave vector $k_x$, $y_0 \in (0,L_y)$. Since $k_x \in(0,L_y/\ell^2)$ the
number of states in a Landau level is given by $N =L_xL_y/2\pi\ell^2$
and the level degeneracy reads
\begin{equation}
\frac{N}{L_xL_y}=\frac{|e|\mathcal{B}}{h}.
\end{equation}

Very similar form of the eigenstates of the full Hamiltonian $H_0$ is
obtained if we sum in Eq.~(\ref{H0}) the terms linear and quadratic in
$y$. Then $H_0$ takes the form
\begin{eqnarray}
\label{H01}
\lefteqn{H_0= \frac{1}{2m}\left(p_x -mv_d-|e|\mathcal{B}\,y\right)^2}
\nonumber \\
 & & \mbox{}+\frac{p_y^2}{2m} + v_d(p_x-mv_d) +\frac{mv_d^2}{2},
\end{eqnarray}
in which the drift velocity $v_d=\mathcal{E}/\mathcal{B}$ of an
electron in crossed fields is employed.

It follows from the similarity between expressions (\ref{H0tilde}) and
(\ref{H01}) that the eigenenergies of $H_0$ can be constructed from
the zero-electric-field ones, $\widetilde{E}_{0n}(k_x)$, given by
Eq.~(\ref{Etilde}). They can be expressed as
\begin{equation}
\label{E0n}
E_{0n}(\bar{k}_x)=\widetilde{E}_{0n}(\bar{k}_x)+\hbar\bar{k}_xv_d 
+\frac{mv_d^2}{2},
\end{equation}
where $\bar{k}_x$ stands for $k_x-mv_d/\hbar$. The Landau levels
become tilted and, therefore, their degeneracy in $k_x$ is lifted. The
eigenfunctions $\varphi_{0n\bar{k}_x}(y)$ remain practically
unchanged, only the wave vector $k_x$ was replaced by $\bar{k}_x$ in
Eq.~(\ref{0phi}).

In comparison with the zero-electric-field case the diagonal matrix
element of the velocity component is nonzero and given by
\begin{equation}
\langle\varphi_{0n\bar{k}_x}|v_x|\varphi_{0n\bar{k}_x}\rangle =
\frac{1}{\hbar}\frac{\partial E_n(\bar{k}_x)}{\partial\bar{k}_x}=v_d,
\end{equation}
and the center of mass of the cyclotron orbit reads
$\langle\varphi_{0n\bar{k}_x}|y|\varphi_{0n\bar{k}_x}\rangle = \ell^2
\bar{k}_x= y_0-v_d/\omega$.

This well known example demonstrates that the electric field does
change the zero-electric-field electronic structure only in an
unessential way.  Both systems are described by almost equivalent
Hamiltonians which differ only by the shift of the coordinate center
and the changes of the constants. We will show in the following
sections that it is valid also in the presence of the superlattice
periodic potential.
\section{The electric field  parallel with the lattice vector
\label{Parallel}}
In the zero-magnetic-field limit this geometrical arrangement would
lead to the standard Bloch oscillations.  In crossed electric and
magnetic fields the Hamiltonian can be written as a sum of $H_0$, as
given by Eq.~(\ref{H01}), and of the  periodic potential  $V(y)$,
\begin{equation}
H = H_0 + V (y).
\end{equation}
The resulting Hamiltonian reads
\begin{eqnarray}
\label{H02}
\lefteqn{H = 
\frac{1}{2m}\left(p_x -mv_d-|e|\mathcal{B}\,y\right)^2 }
\nonumber \\
 & & \mbox{} + \frac{p_y^2}{2m} +V(y)+ v_d(p_x-mv_d) +\frac{mv_d^2}{2}.
\end{eqnarray}
Similarly as in the case of a free electron described above, it is
obvious that the eigenstates of $H$ are closely related to the
eigenstates of the zero-electric-field Hamiltonian
\begin{equation}
\label{H03}
\widetilde{H} = \frac{1}{2m}\left(p_x -|e|\mathcal{B}\,y\right)^2 +
\frac{p_y^2}{2m} +V(y).
\end{equation}
For a given $k_x$ both Hamiltonians reduce to one-dimensional ones, 
which differ only by a constant $v_d\bar{k}_x +{mv_d^2}/{2} $ and
a shift $m v_d/\hbar$ in the wave vector $k_x$. Therefore both 
Hamiltonians yield essentially the same
spectrum of eigenvalues.  

The eigenenergies of the Hamiltonian
(\ref{H02}) can be written, in analogy with  Eq.~(\ref{E0n}), as
\begin{equation}
\label{En}
E_s(\bar{k}_x)=
\widetilde{E}_s(\bar{k}_x)+\hbar\bar{k}_x v_d +\frac{mv_d^2}{2}.
\end{equation}

Note that for the zero-electric-field case the periodic potential
$V(y)$ removes the degeneracy of Landau levels. The
eigenenergies of the Hamiltonian (\ref{H03}) become the Landau
subbands $\widetilde{E}_s(k_x)$, $s=0,1,2,\cdots$, periodic in $k_x$
with the period
\begin{equation}
K=\frac{a}{\ell^{2}}= \frac{a|e|\mathcal{B}}{\hbar}.
\end{equation}
We replace $n$ by the integer $s$ to distinguish between a subband and
a level.  The corresponding eigenfuctions $\widetilde{\varphi}_s(k_x)$
are, similarly as $\varphi_{0nk_x}$, bounded in $y$ direction for any
$k_x$ due to the confinement by the parabolic potential generated by
$\mathcal{B}$.

The diagonal matrix elements of the velocity component $v_x$ in
crossed fields are now given by
\begin{equation}
\langle\widetilde{\varphi}_{s\bar{k}_x}|v_x
|\widetilde{\varphi}_{s\bar{k}_x}\rangle
= \frac{1}{\hbar}\frac{\partial E_s(\bar{k}_x)}{\partial\bar{k}_x}=
\frac{1}{\hbar}\frac{\partial
\widetilde{E}_s(\bar{k}_x)}{\partial\bar{k}_x} +v_d,
\end{equation}
and the center of mass of the cyclotron orbit reads
\begin{equation}
\langle\widetilde{\varphi}_{s\bar{k}_x}|y
|\widetilde{\varphi}_{s\bar{k}_x}\rangle =
 y_0-v_d/\omega - \frac{1}{\hbar\omega}\frac{\partial 
\widetilde{E}_s(\bar{k}_x)}
{\partial\bar{k}_x}.
\end{equation}

It follows from Eq.~(\ref{En}) that $E_s(\bar{k}_x)$ is a
step-like function with the step length $K$ and the step height
$|e|\mathcal{E}a$, i.e. we can write
\begin{equation}
\label{result}
E_s(\bar{k}_x+K) = E_s(\bar{k}_x)+ \hbar \omega_{BO},\,\,\,\,\, 
\omega_{BO}=\frac{|e|\mathcal{E}a}{\hbar},
\end{equation}
where $\omega_{BO}$ denotes the frequency of Bloch oscillations.

This step function replaces the standard Wannier-Stark ladder obtained
in the zero-magnetic-field case. Note that $E_s(\bar{k}_x)$ are the
true eigenenergies and not the resonances as the Wannier-Stark states,
i.e. the magnetic field completely suppresses the Zener tunneling.

The energy spectra of the above Hamiltonians can be found either by
the direct numerical solution of the corresponding one-dimensional
Schr\"odinger equations, or we can look for the eigenfunctions in the
form of the linear combination of the free electron functions
(\ref{0phi}). As the resulting eigenenergies are the functions of
$k_x$, we can evaluate the matrix elements for a given $k_x$ and
diagonalize the resulting matrix.

The matrix elements of the Hamiltonian $\widetilde{H}$,
Eq.~(\ref{H03}), are diagonal in $k_x$.  The periodic potential $V(y)$
can be expanded into the Fourier series,
\begin{equation}
V(y)=\sum_{j}V_j \cos(G_jy),
\end{equation}
with $G_j=jG$, $G=2\pi/a$ being the reciprocal lattice
vector. Therefore, we can look for the eigenfunctions
$\widetilde{\varphi}_{sk_x}(y)$ in the form
\begin{equation}
\widetilde{\varphi}_{sk_x} = \sum_{n'}c_{n'}\psi_{0n'k_x},
\end{equation}
and the matrix elements of $\widetilde{H}$  can be written as
\begin{equation}
\label{without}
\widetilde{H}_{k_x,n,n'} = 
\hbar\omega\left(n+\frac{1}{2}\right)\delta _{n,n'}
+ T_{k_x,n,n'},
\end{equation}
where $T_{k_x,n,n'}$ is the matrix element of a component of the 
periodic potential
\begin{eqnarray}
\lefteqn{T_{k_x,n,n'} =}\nonumber \\
 & & 
\sum_j  V_j \frac{1}{\sqrt{2^n 2^{n'}n!n'!\pi \ell^2}}
\int_{-\infty}^{+\infty}
\exp\left[{-\frac{(y-y_0)^2}{\ell^2}}\right]\nonumber \\
& &
\times H_n\left(\frac{y-y_0}{\ell}\right)
H_{n'}\left(\frac{y-y_0}{\ell}\right)
\cos\left(G_jy\right){\mathrm d}y.
\end{eqnarray}
Introducing the dimensionless variable $\eta=(y-y_0)/\ell$, we 
obtain 
\begin{eqnarray}
\lefteqn{T_{k_x,n,n'} =}\nonumber\\
& &
\sum_j  V_j \frac{1}{\sqrt{2^n 2^{n'}n!n'!\pi}}
\int_{-\infty}^{+\infty}
\exp{\left(-\eta^2\right)}\nonumber \\
 & & 
\times H_n(\eta)H_{n'}(\eta)
\cos(G_j\ell\eta+G_jy_0){\mathrm d}\eta.
\end{eqnarray}
To reduce this expression to a pair of standard tabular
integrals we can  employ the simple trigonometric relation 
\begin{eqnarray}
\label{co}
\lefteqn{cos(G_j\ell\eta+G_jy_0)=}\nonumber\\
& &
\cos(G_j\ell\eta)\cos(G_jy_0)
-\sin(G_j\ell\eta)\sin(G_jy_0).
\end{eqnarray}
After substitution, the examined $T_{k_x,n,n'}$ is divided into two
parts $I_{1,n,n'}$ and $I_{2,n,n'}$. The first part, which includes
the cosine in the integrand, is equal to zero if the integrand is the
odd function. For the even integrand it reads
\begin{eqnarray}
\label{I1_1}
\lefteqn{I_{1,n,n'}=}\nonumber\\
& &
\sum_j  V_j
\frac{2\cos(G_jy_0)}{\sqrt{2^n2^{n'}n!n'!\pi}}\int_{0}^{+\infty}
\exp{\left(-\eta^2\right)}\nonumber \\
 & & 
\times 
H_n(\eta)H_{n'}(\eta)
\cos(G_j\ell\eta){\mathrm d}\eta.
\end{eqnarray}
The explicit analytic form of $I_{1,n,n'}$
(see,e.g.,~\cite{Gradshtejn}) is given by
\begin{eqnarray}
\label{I1_2}
\lefteqn{I_{1,n,n'}=} \nonumber\\
& &
\sum_j  V_j\,
\sqrt{\frac{2^{n}n!}{2^{n'}n'!}}(-1)^{\frac{n'-n}{2}}
(G_j\ell)^{n'-n}\exp\left(-\frac{G^2_j\ell^2}{4}\right)\nonumber \\
 & & 
\times L_n^{n'-n}\left(\frac{G^2_j\ell^2}{2}\right)
\cos(G_jy_0),
\end{eqnarray}
for $ n'-n=2 p$. Here $L_n^{n'-n}$ are the Laguerre polynomials, and $p$ is
an integer number.  The second part $I_{2,n,n'}$ differs from the first one
only by the sine written instead of the cosine
\begin{eqnarray}
\label{I2_1}
\lefteqn{I_{2,n,n'}=}\nonumber\\
& &
-\sum_j  V_j
\frac{2\sin(G_jy_0)}{\sqrt{2^n2^{n'}n!n'!\pi}}\int_{0}^{+\infty}
\exp{\left(-\eta^2\right)}\nonumber \\
 & & 
\times H_n(\eta)H_{n'}(\eta)
\sin(G_j\ell\eta){\mathrm d}\eta.
\end{eqnarray}
The part $I_{2,n,n'}$ is also given by the tabular integrals which are
nonzero for the odd integrands
\begin{eqnarray}
\label{I2_2}
\lefteqn{I_{2,n,n'}=}\nonumber\\
& &
-\sum_j  V_j
\sqrt{\frac{2^{n}n!}{2^{n'}n'!}}(-1)^{\frac{n'-n-1}{2}}
(G_j\ell)^{n'-n-1}\exp\left(-\frac{G^2_j\ell^2}{4}\right)\nonumber \\
& & 
\times L_n^{n'-n}\left(\frac{G^2_j\ell^2}{2}\right)
\sin(G_jy_0),
\end{eqnarray}
for $ n'-n=2p+1$. Note that $k_x$ enters the matrix elements
$T_{k_x,n,n'}$ through the identity $G_j y_0=j 2\pi k_x/K$.  Only the
minor changes are introduced by applying the electric field.  The
matrix elements of the Hamiltonian (\ref{H02}) read
\begin{equation}
\label{with}
H_{\bar{k}_x,n,n'} = 
\left[\hbar\omega\left(n+\frac{1}{2}\right)
+\hbar\bar{k}_{xj}v_d +\frac{mv_d^2}{2}\right]\delta _{n,n'}
+ T_{\bar{k}_x,n,n'}
\end{equation}
The similarity between Hamiltonians (\ref{without}) and (\ref{with})
confirms the validity of the expression (\ref{En}).

As a simple example \cite{Labbe} we present the case of a weak
perturbation $V(y)=V\cos(Gy)$ which does not mix the Landau
levels with different $n$. Then, replacing $n$ by $s$ to stress the
difference between the level and the subband, the eigenenergies for
the zero-electric-field Hamiltonian become
\begin{eqnarray}
\label{zerokx}
\lefteqn{\widetilde{E}_s(k_x)=
\hbar\omega\left(s+\frac{1}{2}\right)}\nonumber\\
& &
+
V \exp\left(-\frac{G^2\ell^2}{4}\right)
L_s\left(\frac{G^2\ell^2}{2}\right)
\cos\left(\frac{2\pi k_x}{K}\right),
\end{eqnarray}
and, finally, we get in accord with Eq.~(\ref{En}),

\begin{eqnarray}
\lefteqn{{E}_s(\bar{k}_x)=
\hbar\omega\left(s+\frac{1}{2}\right)
+\hbar\bar{k}_x v_d +\frac{mv_d^2}{2}
}\nonumber\\
& &
+
V \exp\left(-\frac{G^2\ell^2}{4}\right)
L_s\left(\frac{G^2\ell^2}{2}\right)
\cos\left(\frac{2\pi \bar{k}_x}{K}\right),
\end{eqnarray}
for the eigenvalues in the electric field.
%
\section{The electric field  perpendicular to the lattice vector
\label{perpendicular}}
This arrangement corresponds to the alternative oscillations.  The
Hamiltonian of an electron in crossed electric and magnetic fields
with the periodic modulation in the $x$ direction can be written as a
sum of the Hamiltonian $H_0$, as given by Eq.~(\ref{H01}), and of the
potential $V(x)$,
\begin{equation}
H = H_0 + V (x),
\end{equation}
or, explicitly,
\begin{eqnarray}
\label{H04}
\lefteqn{H = \frac{1}{2m}\left(p_x -mv_d-|e|\mathcal{B}\,y\right)^2}
\nonumber \\
 & & \mbox{}+\frac{p_y^2}{2m} +V(x)+ v_d(p_x-mv_d) +\frac{mv_d^2}{2}.
\end{eqnarray}
We expect that due to the strong confinement by the magnetic field the
eigenfunctions will be localized in $y$ direction.  For
$\mathcal{E}\rightarrow 0$, the eigenenergies should reduce to
$\widetilde{E}_s(k_y)$, periodic in $k_y$ with the period $K$. This
function must be equivalent to $\widetilde{E}_s(k_x)$ found in the
previous section, as the result should not depend on the choice of
calibration of the vector potential.

Since the Hamiltonian (\ref{H04}) cannot be simply reduced to the
one-dimensional one, we will look for the eigensolutions starting from
the linear combination of eigenfunctions of $H_0$.  The potential
$V(x)$ will be, similarly as in Sec. \ref{Parallel}, considered in the
form of the Fourier series
\begin{equation}
V (x) =\sum_j
V_j \cos(G_jx).
\end{equation}
The functions $\psi_{s\bar{k}_x}(\vec{r})$, we are looking for, can be
written as a Bloch sum,
\begin{equation}
\psi_{s\bar{k}_x}(\vec{r}) = 
\sum_{n'j'}c_{n'j'}{\mathrm e}^{i\bar{k}_{xj'}x}
\varphi_{0n'\bar{k}_{xj'}}(y),
\end{equation}
where $\bar{k}_{xj}= \bar{k}_x+G_j$, i.e. as the linear combination of
functions the centers of which, $\ell^2\bar{k}_{xj}$, are arranged
periodically along the $y$ axis with the distance $\ell^2G$.

For a given $\bar{k}_x$, the matrix elements of the Hamiltonian
(\ref{H04}) can be written as
\begin{eqnarray}
\label{hwith}
\lefteqn{H_{\bar{k}_x,nj,n'j'}=}\\ 
\nonumber
& &\left[\hbar\omega\left(n+\frac{1}{2}\right)
+\hbar\bar{k}_{xj'}v_d +\frac{mv_d^2}{2}\right]\delta_{nj,n'j'}
 + T_{nj,n'j'}.
\end{eqnarray}
Here $T_{nj,n'j'}$ denotes the matrix elements of the potential $V(x)$.
They are products of the matrix elements of $V_j\cos(G_jx)$, calculated
between the plane-wave parts of the wave functions, and of the overlap
integrals of localized functions $\varphi_{0n'\bar{k}_{xj'}}(y)$. Only
the matrix elements $V_{j-j'}$ are nonzero. The centers of
overlapping functions are $\ell^2\bar{k}_{xj}$ and
$\ell^2\bar{k}_{xj'}$, respectively, and their distance is
$\ell^2G_{j-j'}$. Then $ T_{nj,n'j'}$ are equal to
\begin{eqnarray}
\lefteqn{T_{nj,n'j'} =  V_{j-j'}\sqrt{\frac{2^{n'}n!}{2^{n}n'!}}
\exp\left(-\frac{ G^2_{j-j'}\ell^2}{4}\right)} \nonumber \\
& &
\times \left(\frac{ G_{j-j'}\ell}{2} \right)^{n'-n}
L_n^{n'-n}\left(\frac {G^2_{j-j'}\ell^2}{2}\right)
\end{eqnarray}
for $n'\ge n$ and
\begin{eqnarray}
\lefteqn{T_{nj,n'j'} =  V_{j-j'}
\sqrt{\frac{2^{n}n'!}{2^{n'}n!}}
\exp\left(-\frac{ G^2_{j-j'}\ell^2}{4}\right)}\nonumber \\
& &\times  \left(-\frac{ G_{j-j'}\ell}{2} \right)^{n-n'}
L_{n'}^{n-n'}\left(\frac{ G^2_{j-j'}\ell^2}{2}\right)
\end{eqnarray}
for $n'\le n$. They decrease exponentially with the distance between the 
centers of functions  $\varphi_{0n\bar{k}_{xj}}(y)$ and 
$\varphi_{0n'\bar{k}_{xj'}}(y)$.

Note that the electric field $\mathcal{E}$ (the drift velocity $v_d$)
and the wave vector $\bar{k}_x$ enter only  the diagonal part of the
Hamiltonian, and, therefore, only the diagonal elements of the
corresponding matrix equation which reads
\begin{equation}
\label{schwith}
\sum_{n'j'}\left(H_{\bar{k}_x,nj,n'j'}-E\right)  c_{n'j'}= 0.
\end{equation}
This allows us to predict an interesting $\bar{k}_x$ dependence of the
eigenstates of the above equation. 

First, it follows from the form of the matrix elements of the
Hamiltonian~(\ref{hwith}) that the eigenenergies depend linearly on
$\bar{k}_x$, and that the electrons move, in spite of the presence of
the superlattice potential, with the same drift velocity $v_d$ as 
free electrons in crossed fields. Moreover, if we replace $\bar{k}_x$
by $\bar{k}_x+G$, we can rewrite the Hamiltonian matrix elements
(\ref{hwith}) to the form
\begin{equation}
\label{hwithout}
H_{\bar{k}_x+G,nj,nj}= H_{\bar{k}_x,nj,nj} +\hbar G v_d.
\end{equation}
Then equation 
\begin{equation}
\label{schwith+G}
\sum_{n'j'}\left(H_{\bar{k}_x+G,nj,n'j'}-E\right)  c_{n'j'}= 0
\end{equation}
can be replaced by
\begin{equation}
\label{schwith-G}
\sum_{n'j'}\left(H_{\bar{k}_x,nj,n'j'}-E'\right)  c_{n'j'}= 0
\end{equation}
where $E'$ stands for $E-\hbar G v_d$. Therefore we can conclude for
the corresponding eigenvalues that
\begin{equation}
E_s(k_x+G) = E_s(k_x)+ \hbar\omega_{\mathcal{B_\parallel}}, \,\,
\omega_{\mathcal{B_\parallel}}=\frac{2\pi v_d}{a}.
\end{equation}
The eigenvalues of Eq.~(\ref{schwith}) should reduce for $\mathcal{E}
\rightarrow 0$ to the zero-field eigenvalues $\widetilde{E}_s$
periodic in $k_y$.

We will again illustrate the above consideration on the simple case of
a weak perturbation $V \cos(Gx)$.

We start with the case of the zero electric field \cite{Labbe} to show
how the $k_y$ dependence of $\widetilde{E}_s$ is introduced.

The function
\begin{equation}
\label{psi}
\widetilde{\psi}_{sk_x}(\vec{r}) =
\sum_{j'}c_{j'}{\mathrm e}^{ik_{xj'}x}\varphi_{0n k_{xj'}}(y) =
\sum_{j'}c_{j'}|n j'\rangle
\end{equation}
is assumed not to mix the Landau levels. Note that $k_x\in(-\pi/a,\pi/a)$
and $j'=1,2,\cdots,j_M$ where $j_M=L_ya/2\pi\ell^2$.
The equation
\begin{equation}
\label{baf}
\langle nj|H - E |\widetilde{\psi}_{sk_x}\rangle = 0
\end{equation}
implies
\begin{eqnarray}
\lefteqn{c_j\left[\hbar\omega\left(n+\frac{1}{2}\right)-E\right] }\\ 
\nonumber
& &  
+\frac{V}{2} \exp\left(-\frac{G^2\ell^2}{4}\right)
L_n\left(\frac{G^2\ell^2}{2}\right)
\left(c_{j-1} + c_{j+1}\right)= 0.
\end{eqnarray}
We write $c_j$ in the form $c_j=c_0\exp(iqj)$
and look for $c_0$ and $q$ instead of $c_j$. The first consequence is that
the solution of Eq.~(\ref{baf}) can be written in the form
\begin{equation}
\widetilde{E}_s = \hbar\omega\left(n+\frac{1}{2}\right) 
+ V\exp\left(-\frac{G^2\ell^2}{4}\right)
L_n\left(\frac{G^2\ell^2}{2}\right)
\cos q.
\end{equation}
It follows from
$\langle\widetilde{\psi}_{sk_x}|\widetilde{\psi}_{sk_x}\rangle =1$
that $j_Mc_0^2=1$ and $c_0^2=\ell^2G/L_y$.

If we write $\widetilde{\psi}_{sk_x,q}$ as
\begin{equation}
\widetilde{\psi}_{sk_x,q} =
c_0 
\sum_{j'}{\mathrm e}^{iqj'}|n j'\rangle
\end{equation}
we get from
\begin{equation}
\langle\widetilde{\psi}_{sk_x,q}|\widetilde{\psi}_{sk_x,q'}\rangle = 0
\end{equation}
the expression $q=(2\pi\nu/L_y)\ell^2G, \nu=1,2,\cdots$. Taking into
account that $2\pi\nu/L_y=k_y$ and $x_0=\ell^2 k_y$, we arrive at
$q=x_0G=2\pi k_y/K$ and finally
\begin{eqnarray}
\lefteqn{\widetilde{E}_s(k_y)=
\hbar\omega\left(s+\frac{1}{2}\right)}\nonumber\\
& &
+
V \exp\left(-\frac{G^2\ell^2}{4}\right)
L_s\left(\frac{G^2\ell^2}{2}\right)
\cos\left(\frac{2\pi k_y}{K}\right).
\end{eqnarray}
As expected, the change of the direction of the lattice vector has no
influence and the only difference in comparison with the $V(y)$ case, see
Eq.~(\ref{zerokx}), is replacement of $k_x$ by $k_y$.

It is not too difficult to introduce the electric field into this
simple model.  First, the $k_x$ must be replaced by $\bar{k}_x$ and
$q$ by $k_y=K q/2\pi$ in $\widetilde{\psi}_{s{k}_x,q}$. Then we can
calculate the matrix elements of the Hamiltonian~(\ref{hwith}), which
include the $\bar{k}_x$ and $v_d$ dependent parts. We get
\begin{eqnarray}
\lefteqn{\langle\widetilde{\psi}_{s\bar{k}_x,q}|H|
\widetilde{\psi}_{s\bar{k}_x,q'}\rangle = }\nonumber\\
& &
\delta_{k_y,k'_y}\left[\widetilde{E}_s(k_y)
+\hbar \bar{k}_x v_d +\frac{mv_d^2}{2} 
-i|e|\mathcal{E}\frac{d}{dk_y}\right],
\end{eqnarray} 
and thus, taking into account the periodicity in $k_y$, the eigenenergies
of the corresponding equation can be easily calculated:
\begin{eqnarray}
\lefteqn{E_{r,s,\bar{k}_x} =
\frac{1}{K}\int_0^K\widetilde{E}_s(k_y)dk_y }\nonumber\\
&& 
+\hbar \bar{k}_x v_d +\frac{mv_d^2}{2} 
+ r \hbar \omega_{\mathcal{B_\parallel}}, \,\,\, r=0, \pm 1,\cdots.
\end{eqnarray}
In this example we limited our consideration to one subband
$\widetilde{E}_s(k_y)$.  Note that for more subbands the electric
field cannot cause the intersubband transition, as we use the
electric-field-dependent basis of functions ${\mathrm
e}^{i\bar{k}_{xj'}x} \varphi_{0n'\bar{k}_{xj'}}(y)$. Therefore, the
solutions of Eq.~(\ref{schwith}) are the eigenstates and not the
resonances.
\section{Conclusions}
The presented theoretical analysis has implicitly assumed a single
electron model for a Bloch electron in crossed electric and magnetic
fields, $\vec{\mathcal{E}}\perp\vec{\mathcal{B}}$, moving in a perfect
superlattice crystal with the lattice vector $\vec{a}$.  Two
geometrical arrangements were considered.

In the case $\vec{a}\parallel\vec{\mathcal{E}}$, which would
correspond to the Bloch oscillations for $\mathcal{B}=0$, the magnetic
field converts the standard Wannier-Stark ladder of resonances to the
step-like eigenenergies, which are functions of the wave vector
$\vec{k}_{\perp}$, perpendicular to both $\vec{\mathcal{E}}$ and
$\vec{\mathcal{B}}$. The step length is $a|e|\mathcal{B}/\hbar$. The
step height $\hbar \omega_{BO} $ corresponds to the frequency of the
Bloch oscillations $\omega_{BO}=|e|\mathcal{E}a/\hbar$.

The alternative magnetic-field-induced oscillations were suggested as
a possible source of  terahertz radiation~\cite{or} for the case
$\vec{a}\perp \vec{\mathcal{E}}$.  The electron motion is composed
from oscillations along $\vec{a}$ and the drift due to the Lorentz
force with the velocity $v_d=\mathcal{E}/\mathcal{B}$ in the direction
of $\vec{k}_{\perp}$.  The resulting eigenstates resemble those of a
free electron in crossed fields. The eigenenergies $E_M$ depends
linearly on the $k_{\perp}$, $E_M \propto \hbar v_d k_{\perp}$, their
separation on the energy scale is
$\hbar\omega_{\mathcal{B_{\parallel}}}$.
 
The above conclusions are valid for the optimal conditions for
coherent Bloch and alternative oscillations, neglecting the importance
of additional scattering which can lead to damping of oscillations.
\section{Acknowledgements}
This work has been supported by the Ministry of Education of the Czech
Republic Center for Fundamental Research LC510, the Ministry of Education 
of the Czech Republic research plan MSM 0021620834, and Academy of Sciences
of the Czech Republic project KAN400100652.
%


\begin{thebibliography}{99}

\bibitem{cap}
J.\  Faist, F.\ Capasso, D.\ L.\ Sivco, C.\ Sirtori, A.\ L.\
Hutchinson and A.\ Y.\ Cho, 
 Science {\bf 264},  553 (1994).

\bibitem{Feldman}
J.\ Feldmann, K.\ Leo, J.\ Shah, D.\ A.\ B. Miller, J.\ E.\ Cunningham,
T.\ Meier, G.\ von Plessen, A.\ Schulze, P.\ Thomas, and S.\ Schmitt-Rink,
Phys.\ Rev.\ B {\bf 46}, 7252 (1992).  

\bibitem{Waschke}
C.\ Waschke, H.\ G.\ Roskos, R.\ Schwedler, K.\ Leo, H.\ Kurz, and
K.\ K\"{o}hler, Phys.\ Rev.\ Lett.\ {\bf 70}, 3319 (1993).

\bibitem{Deko}
T.\ Dekorsy, P.\ Leisching, K.\ K\"{o}hler, and H. Kurz,
Phys.\ Rev.\ B {\bf 50}, 8106 (1994).
 
\bibitem{Cho}
G.\ C.\ Cho, T.\ Dekorsy, H.\ J.\ Bakker, H.\ Kurz, A.\ Kohl, and B.\ Opitz,
Phys.\ Rev.\ B {\bf 54} 4420 (1996).

\bibitem{Lys}
V.\ G.\ Lyssenko, G.\ Valusis, F.\ L\"{o}ser, T.\ Hasche, K.\ Leo,
M.\ M.\ Dignam, and K.\ K\"{o}hler,
Phys.\ Rev.\ Lett.\ {\bf 79}, 301 (1997).

\bibitem{Hart}  
T.\ Hartmann, F.\ Keck, H.\ J.\ Korsch, and S.\ Mossmann, 
New J.\ Phys. {\bf 6}, 2 (2004).


\bibitem{Leo}
K.\ Leo, Semicond.\ Sci.\ Technol.\ {\bf 13} 249 (1998).

\bibitem{Patane}
A.\ Patan\`e, N.\ Mori, D.\ Fowler, L.\ Eaves, M.\ Henini, D.\ K.\ Maude,
C.\ Hamaguchi, and R.\ Airey,
Phys.\ Rev.\ Lett.\ {\bf 93}, 146801  (2004). 

\bibitem{Scalari} 
G.\ Scalari, S.\ Blaser,J.\ Faist, H.\ Beere, E.\ Linfield, D.\ Ritchie,
and G.\ Davies,
Phys.\ Rev.\ Lett.\ {\bf 93}, 237403 (2004).


\bibitem{or}
 M.\ Orlita, R.\ Grill, L.\ Smr\v{c}ka, and M.\ Zv\'{a}ra, Phys.\ Rev.\
 B  {\bf 74}, 125312 (2006).

\bibitem{qureshi}
 N.\ Qureshi, ``Terahertz Dynamics of a Superlattice in Crossed
 Electric and Magnetic Fields'' Ph.D. thesis, University of
 California, Santa Barbara, 2002.

\bibitem{Wong}
C.\ Wang and J.\ C.\ Cao, Phys.\ Rev.\ B {\bf 72}, 045339 (2005). 

\bibitem{From1}
T.\ M.\ Fromhold, A.\ A.\ Krokhin, C.\ R.\ Tench, S.\ Bujkiewicz,
P.\ B.\ Wilkinson, F.\ W.\ Sheard, and L. Eaves, Phys.\ Rev.\ Lett.\ 
{\bf 87}, 046803 (2001).

\bibitem{From2}
T.\ M.\ Fromhold, A.\ Patan\`e, S.\ Bujkewicz, P.\ B.\ Wilkinson, 
D.\ Fowler,
D.\ Sherwood, S.\ P.\ Stapleton, A.\ A.\ Krokhin, L.\ Eaves, M.\ Henini,
N.\ S.\ Sankeshwar, and F.\ W.\ Sheard, Nature (London) 428, 726 (2004).

\bibitem{ne}
G.\ Nenciu, Rev.\ Mod.\ Phys.\ {\bf 63}, 91 (1991).


\bibitem{so}
V.\ N.\ Sokolov, L.\ Zhou, G.\ J.\ Iafrate, and J.\ B.\ Krieger,
 Phys.\ Rev.\ B  {\bf 73}, 205304 (2006).

\bibitem{ya06}
Lijun Yang and Marc M.\ Dignam,  Phys.\ Rev.\ B  {\bf 73},
075319 (2006).

\bibitem{ya05} 
 Lijun Yang, Ben Rosam, K.\ Leo, and Marc M.\ Dignam, Phys.\ Rev.\ B  
{\bf 72}, 115313 (2005).

\bibitem{davies} J.\ H.\ Davies, {\it The~Physics~of~Low-Dimensional
Semiconductors: An Introduction} (Cambridge University Press, Cambridge
1997), p.\ 229.

\bibitem{Gradshtejn}
I.\ S.\ Gradshteyn, I.\ M.\ Ryzhik, {\it Tables of integrals, sums,
series, and products} (Moscow, 1963), p. 852-855. 

\bibitem{Labbe}
J.\ Labb\'e,  Phys.\ Rev.\ B  {\bf 35}, 1373 (1987).

\end{thebibliography}
\end{document}